# Defect-Engineered Beryllium Dinitride ($BeN_2$) Monolayer with Light-Metal Decoration for Reversible High-Capacity Hydrogen Storage

Wael Othman[1,2,¶], Ibrahim Alghoul[3,4,¶], K-F. Aguey-Zinsou[5], Nacir Tit[3,4*], and Tanveer Hussain[6*]
[1] Biomedical Engineering and Biotechnology, Khalifa University, Abu Dhabi, United Arab Emirates
[2] Healthcare Engineering Innovation Group (HEIG), Khalifa University, Abu Dhabi, United Arab Emirates
[3] Physics Department, United Arab Emirates University, Al Ain, United Arab Emirates
[4] Water and Energy Research Center, United Arab Emirates University, Al Ain, United Arab Emirates
[5] MERLin, School of Chemistry, University of Sydney, NSW 2006, Australia
[6] School of Science and Technology, University of New England, Armidale, New South Wales, 2351, Australia
* Correspondence: ntit@uaeu.ac.ae & tanveer.hussain@une.edu.au
¶ These two authors have equal contributions

**Abstract:**
Hydrogen ($H_2$) possesses the highest gravimetric energy density of any chemical fuel and is the most abundant element in the universe; however, its extremely low volumetric energy density at standard conditions imposes a fundamental materials challenge for safe, efficient, and reversible storage. Here, we report a defect-engineered two-dimensional beryllium dinitride ($BeN_2$) monolayer that enables stable light-metal functionalization for high-capacity $H_2$ storage. A 2×2 supercell containing two intrinsic beryllium vacancies accommodates four Li, Na, and K atoms without clustering, exhibiting strong average metal-vacancy binding energies of -3.80, -2.94, and -3.18 eV, respectively. Ab initio molecular dynamics simulations at 400 K confirm the thermal stability of the metal-decorated frameworks and the suppression of metal aggregation. The vacancy-stabilized alkali-metal centers generate localized charge polarization that facilitates the adsorption of up to $20H_2$ molecules per supercell, with average adsorption energies of −0.182 eV (Li), −0.191 eV (Na), and −0.171 eV (K), making the adsorption reversible under near-ambient conditions. The corresponding gravimetric $H_2$ storage capacities reach 11.64, 9.82, and 8.49 wt%, respectively, significantly exceeding the US Department of Energy (DOE) ultimate target of 6.50 wt%. Moreover, thermodynamic analysis further confirms favorable adsorption-desorption behavior within practical operating windows. These results establish vacancy-defected light-metal decoration $BeN_2$ as a viable design strategy for high-density, reversible $H_2$ storage, providing a scalable framework for engineering polar lightweight materials for energy storage applications.

## I. Introduction

The transition toward a hydrogen-based economy marks a critical pivot in global energy strategy, aiming to balance the rising demands of a growing population with the urgent need to reduce environmental damage from fossil fuel combustion. Hydrogen ($H_2$) is highly attractive due to its abundance in nature and its low-emission production through either the catalytic methane reforming or water electrolysis. Despite these advantages, its physical properties pose significant engineering challenges. In its gaseous form, $H_2$ has extremely low volumetric energy density, while conventional storage via cryogenic liquefaction at temperatures near 20 K (-253 °C) is highly energy-intensive, limiting its practicality for widespread mobile applications [1]. Further, $H_2$ storage via high-pressure gas compression incurs substantial parasitic energy costs, operational complexity, and inherent safety risks.

To overcome these thermodynamic limitations, research has increasingly shifted toward solid-state $H_2$ storage, particularly through intercalation in large-surface-area van der Waals (vdW) materials. By exploiting the narrow interlayer spaces of two-dimensional (2D) materials such as graphene and transition metal dichalcogenides (TMDs), researchers aim to achieve high $H_2$ uptake capacity via strong physisorption or weak chemisorption. Central to this effort is the ability to tune the adsorption energy of these materials, enabling $H_2$ to be stored and released under near-ambient conditions. Achieving this would meet stringent gravimetric and volumetric targets, bringing $H_2$ closer to viability as a sustainable fuel in fuel cell engines to replace internal combustion engines [2].

The breakthrough invention of graphene in 2004 by Geim and Novoselov has paved the way for a new era of 2D materials physics, enriched with many other exciting discoveries and inventions [3-6]. Despite the extensive expansion of the 2D-material family, such as boron nitride [7], transition-metal dichalcogenides (TMDs) [8], ZnO [9-12], h-BN [13], and MXenes [14], the monolayers that combine high carrier mobility, moderate direct band gap, and outstanding stability are still scarce. For instance, graphene is famous for its ultrahigh mechanical strength and excellent carrier mobility (exceeding $10^5$ $cm^2.V^{-1}.\ s^{-1}$ due to its unique 2D massless Dirac fermions); however, it lacks an intrinsic band gap, limiting its applications in field-effect transistors (FETs) [15-16]. On the other hand, in h-BN and TMDs (e.g., $MoX_2$ and $WX_2$, with X = S, Se), it has been well established that the monolayers have a direct band gap at the high-symmetry point “K”, whereas the multilayer structures become indirect semiconductors [17-18]. Hence, layer stacking can drastically degrade the optical properties.

Searching for 2D multifunctional polynitride materials with novel properties and practical applications remains an attractive challenge. Recently, using the global structure search method combined with the first-principles theory, Zhang and Sun [19] predicted the beryllium dinitride flat monolayer (α-2D-$BeN_2$ hereafter denoted as $BeN_2$). The $BeN_2$ is composed of penta-, hexa-, and hepta-atomic rings and an $N_4$ tetramer in its planar anisotropic structure. $BeN_2$ exhibited high lattice dynamic stability, excellent thermal stability, a moderate bandgap ($E_g$ = 2.23-2.26 eV), high carrier mobilities, high visible-light absorption, and an outstanding potassium (K) storage capacity (~ 2,895 mAh. $g^{-1}$) [20]. It was further showcased that $BeN_2$ is a multifunctional material with great potential for use as a visible-light detector, ductile material, iontronic device, and potassium-ion anode material. It is worth emphasizing that the inventors of $BeN_2$ [19] have demonstrated its possession of three very important characteristics or advantages: (1) The direct bandgap feature of $BeN_2$ sheet is insensitive to layer-stacking pattern and layer number [19, 21]. (2) When rolled up, all the resulting $BeN_2$ nanotubes have a direct bandgap independent of chirality and diameter. (3) The intrinsic acoustic-phonon-limited carrier mobility of $BeN_2$ can reach $\mu_e \sim 10^5$ $cm^2.V^{-1}.\ s^{-1}$ for electron and $10^4$ $cm^2.V^{-1}.s^{-1}$ for hole, which are higher than those of $MoS_2$ and black phosphorous [19]. Furthermore, Mohebbi et al. [22] used density functional theory (DFT) to demonstrate the dynamic stability of $BeN_2$ by showing that its phonon spectrum contains only real modes. The DFT simulations reported by Kishore et al. [23] using VASP have further confirmed the reliability of $BeN_2$ for both oxygen evolution reaction (OER) and $H_2$ evolution reaction (HER), and consequently its suitability as a photocatalyst for water splitting. Finally, Zhou et al. [24] used the atomistic toolkit package to demonstrate the relevance of $BeN_2$ to field-effect transistor performance.

Based on the above-described characteristics of $BeN_2$, as reported in the literature [19-24], we aim in the present investigation to assess its relevance for $H_2$ storage. The defect engineering method is used to modulate $H_2$ adsorption properties, including Be-vacancy formation and surface functionalization, like what we recently reported for defected $BeN_2$ [25]. We plan to explore the functionalization of Be-vacancy-defected $BeN_2$ with light-meta-atoms (i.e., alkali atoms of Li, Na, and K). We will show that the metal-doped $BeN_2$ is thermally stable using ab initio molecular dynamics (AIMD) at 400 K, and that the surface becomes highly polar, capable of inducing electric dipoles in nearby $H_2$ molecules. The paper is organized as follows. The next section describes the computational methods, which are DFT, AIMD, and thermodynamic analysis. In section 3, we discuss the results. Section 4 summarizes our main findings and conclusions.

## II. Computational Methods

Density functional theory (DFT) calculations were performed using the Vienna Ab-initio Simulation Package (VASP) [26]. In the present scheme, the atomic orbitals are expanded in projected augmented waves (PAW) with an energy cutoff of 500 eV [26]. The exchange-correlation potential was described by the general gradient approximation (GGA) within the Perdew-Burke-Ernzerhof (PBE) hybrid functional [27]. To account for van der Waals interactions, we used the DFT-D3 method of Grimme [28]. The modeled structures throughout this study were based on a supercell composed of 2×2 primitive cells of $BeN_2$ having two Be vacancies ($BeN_2:2V_{Be}$), which belongs to a 2D triangular Bravais lattice with lattice constant a = b = 16.038 Å, and composed of 24 atoms (18 C + 6 N) as illustrated in Figure 1a. Adequate vacuum space of 15 Å is added to the vertical direction to ensure the isolation of the $BeN_2:2V_{Be}$ monolayer and its mirror images formed by the implementation of the periodic boundary conditions. For the Brillouin-zone sampling, we employed the Monkhorst-Pack technique [29] with a k-mesh of 15×15×1. In atomic relaxation, we used a convergence tolerance of $10^{-6}$ eV for energy and 0.01 eV/Å for atomic forces. To study the charge exchange, both Bader charge analysis [30] and charge density difference (CDD) [31] were explored.

The average binding energy ($E_{bind}$) of the light-metal (M) atom embedded in the pores of $BeN_2:2V_{Be}$ monolayer is calculated using the formula:

$$E_{bind} = \frac{E_{BeN2:VBe-nM} - E_{BeN2:VBe} - nE_M}{n} \quad (1),$$

where $E_{BeN2:VBe+nM}$, $E_{BeN2:VBe}$, and $E_M$ stand for the total energies of the system of $BeN_2:2V_{Be}$ monolayer with and without the embedded “n” metal “M” atoms, and the isolated metal “M” atom, respectively. The $E_{bind}$ of metal atoms will be compared to the cohesive energies to assess their embedding stability. Additionally, ab initio molecular dynamics (AIMD) simulations at 400 K will be performed on the embedded structures to confirm thermodynamic stability. The Nose-Hoover thermostat is to be employed to control the temperature, with a time step of 1.0 fs over a period of 5.0 ps.

The average adsorption energy ($E_{ads}$) of $H_2$ molecules on top of the substrate is defined as:

$$E_{ads} = \frac{E_{Sheet+mH_2} - (E_{Sheet} + mE_{H_2})}{m} \quad (2),$$

where, $E_{Sheet+mH_2}$, $E_{Sheet}$, and $E_{H_2}$ are the total energies of the $m$$H_2$ molecules adsorbed on the substrate of $BeN_2:2V_{Be}$, the free substrate, and the isolated $H_2$ molecule, respectively. The uptake gravimetric capacity is defined as [32]:

$$C_T(wt\%) = \left[\frac{N_T \cdot M(\mathrm{H}_2)}{N_T \cdot M(\mathrm{H}_2) + M(Sub)}\right] \times 100\% \quad (3),$$

where $M(\mathrm{H}_2)$ and $M(\mathrm{Sub})$ stands for the atomic masses of the $H_2$ molecule and the substrate, whereas $N_T$ stands for the theoretically calculated number of adsorbed $H_2$ molecules. The substrate’s composition is $4M@BeN_2:2V_{Be}$, where M = Li, Na, K.

The adsorption-desorption characteristics at operational conditions were analyzed by statistical thermodynamics according to the grand canonical partition function ($z$):

$$z = 1 + \sum_{i=1}^{n} e^{-\frac{(E_b^i - \mu)}{k_B T}} \quad (4),$$

where $n$ represents the maximum number of adsorbed $H_2$ molecules, whereas $E_b^i$, $k_B$, $\mu$, and $T$ represent the adsorption energy of the $i^{th}$ adsorbed $H_2$ molecule, the Boltzmann constant (1.38 x $10^{-23}$ J/K), the chemical potential of the gas phase of the $H_2$ molecules, and the absolute temperature, respectively. In particular, $\mu$ is a function of $P$ and $T$, defined as:

$$\mu_{H_2}(P,T) = \Delta H + T\Delta S + k_B T ln\frac{P}{P_0} \quad (5),$$

where $\Delta H$, $\Delta S$, $P$, and $P_0$ represent the enthalpy change, entropy change, pressure, and the atmospheric pressure (1.01 x $10^5$ Pa), respectively. The values of $\Delta H$ + $T\Delta S$ are obtained from the experimental database [33]. Meanwhile, the number of stored $H_2$ molecules ($N$) in the host material can be expressed by:

$$N = N_0 \left[\frac{Z-1}{Z}\right] \quad (6),$$

where $N_0$ represents the maximum number of $H_2$ molecules adsorbed on the host medium (namely, N =$N_T$ at 0 K).

# III. Results and Discussion

## a. Structural and Electronic Properties of $BeN_2$

Figure 1a(i) shows the atomically relaxed structure of pristine $BeN_2$, which belongs to the family of Bravais 2D triangular lattice, of point-group symmetry PMC21 (space group No: 26) [20], with a basis of two weights of $BeN_2$ (i.e., 2Be and 4N atoms) per primitive cell. The lattice constant is a = b = 4.541 Å, which is consistent with the value of 4.540 Å reported by Mohebbi et al. [22]. Besides, the bond lengths d(N-N) = 1.32 Å, d(Be-N) = 1.61 Å, and bond angles θ(Be-N-Be) = 109 °, θ(Be-N-N) = 126 ° are in excellent agreement with those ab-initio results reported in literature [22-23]. Both sub-lattices of Be and N atoms lie in the same plane with complete absence of buckling as shown by the side view in Figure 1a(i). The spin-polarized band structure and both projected and total densities of states are shown in Figure 1b(i). As mentioned in the previous section, the GGA-PBE level of theory should be sufficient for our study of $H_2$ adsorption. In the case of pristine $BeN_2$, the band structure shows that it has a direct bandgap at the center of the Brillouin zone (Γ-point) of value: $E_g$ = 1.34 eV, which is in good agreement with the results of 1.32 eV obtained by VASP using the PBE level of theory [22]. The Fermi level is taken as an energy reference ($E_F$=0). Furthermore, Figure 1b(i) shows that the system is behaving as a paramagnetic semiconductor, as there is no distinction between spin-up and spin-down states in both band and DOS profiles. The average cohesive energy of pristine $BeN_2$ obtained from the total energy calculation is: $E_{coh}$= 5.03 eV/atom, which is clearly larger than silicene (4.57 eV/atom) [23], phosphorene (3.48 eV/atom) [23], and $Al_2C$ sheet (4.58 eV/atom) [23], reveals the strong lattice stability of $BeN_2$ structure.

**b. Vacancy-Defected $BeN_2$ ($BeN_2$:$2V_{Be}$)**

Figure 1a(ii) shows the atomically relaxed structure of a 2D supercell of $BeN_2$:$2V_{Be}$, which maintains its constituent atoms within the same membrane, as shown in the side view. $BeN_2$:$2V_{Be}$ retained symmetry properties, such as the diagonal mirror and the Bravais lattice type, remaining triangular but with a larger lattice constant, a = b = 9.081 Å. The stability of this structure will be assessed later after functionalization with alkali metal atoms. Figure 1b (ii) shows the spin-polarized band structure and TDOS/PDOS. The Fermi level is taken as an energy reference ($E_F$=0). The band structure shows that the system behaves as a metal, as two spinless dispersive bands cross the Fermi level at the K point. As far as the magnetic properties are concerned, the system behaves as a paramagnetic metal with no distinction between spin-up and spin-down states, as illustrated in both the band structure and TDOS/PDOS profiles. Moreover, the multiple picks appearing in the TDOS/PDOS structures should be attributed to the localized states in the dangling bonds of N atoms of dual coordination (i.e., poorly coordinated by having two neighbors).

**c. Light Metal Functionalization**

As an initial step, we examined the adsorption behavior of a single $H_2$ molecule on the surface of $BeN_2$:$2V_{Be}$. Figure S1(a) shows two inequivalent sites at the centers of the big pore and the $B_2N_6$ hexagon (X1 and X2), respectively. We place the $H_2$ molecule at a randomly selected vertical height of 1.50 Å and perform structural relaxation. The corresponding $E_{ads}$ values, as shown in Figure S1(b), are -0.156 eV and -0.205 eV, respectively. The recommended average $E_{ads}$ per $H_2$ molecule should be within the -0.20 to -0.60 eV. Although Figure S1(b) shows reasonable $E_{ads}$, they are insufficient because successive $H_2$ adsorption reduces the average binding strength. To address this limitation, we functionalize the system with light alkali metals to enhance adsorption performance.

We functionalized $BeN_2$:$2V_{Be}$ by sequentially decorating the vacancy sites with selected light metal dopants (M = Li, Na, K). At each step, atomic relaxation was performed and the total energy recorded to evaluate the $E_{bind}$. Relaxed structures for 1M@$BeN_2$:$2V_{Be}$, 2M@$BeN_2$:$2V_{Be}$, and 3M@$BeN_2$:$2V_{Be}$ are presented in Figure S2, whereas 4M@$BeN_2$:$2V_{Be}$ are shown in Figure 2. The top views display structures at 0 K, while the two side views show configurations at 0 K and 400 K, respectively. Figure 3a shows the absolute value of the average $E_{bind}$ of 1M, 2M, 3M, and 4M decorated $BeN_2$:$2V_{Be}$. Two key trends emerge: (i) The average $E_{bind}$ decreases with successive metal decoration, which is expected since each subsequent atom occupies progressively less favourable sites. (ii) All average $E_{bind}$ values are stronger than their corresponding bulk cohesive energies $|E_{bind}| > |E_{coh}|$: 1.63 eV for Li, 1.11 eV for Na, and 0.93 eV for K, indicating that the dopants preferentially bind to the lattice rather than cluster.

Thermal stability is crucial for practical $H_2$ storage applications, which operate within the 0-100 °C range and require robust material performance during repeated adsorption/desorption cycles. To validate thermal stability, we performed AIMD simulations at 400 K over 6 ps (Figure 3b). All structures remained intact throughout the simulation, with total energy fluctuations confined below 0.50 eV (<1%), confirming the thermodynamic stability of the 4M@$BeN_2$:2$V_{Be}$ configurations. These negligible energy fluctuations indicate that the decorated monolayers maintain structural integrity under thermal stress, supporting their viability for practical hydrogen storage applications

Figure 4 shows the spin-polarized band structures and related PDOS/TDOS of 4M@$BeN_2$:2$V_{Be}$ monolayers. The Fermi level is taken as an energy reference ($E_F$=0). The common trend among these three structures is that they are paramagnetic semiconductors with band gaps of about 0.64, 0.97, and 0.90 eV, for 4Li@$BeN_2$:2$V_{Be}$, 4Na@$BeN_2$:2$V_{Be}$, and 4K@$BeN_2$:2$V_{Be}$, respectively. It is interesting to see the lattice restoring its original semiconducting character after embedding the metal dopants. Apparently, the metal dopants played a role in passivating the dangling bonds established by Be vacancies. Table 1 summarizes the results of the electronic properties for pristine $BeN_2$, $BeN_2$:2$V_{Be}$ before and after metal functionalization, which correspond to the three configurations illustrated in Figure 4. Remarkably, $BeN_2$:2$V_{Be}$ exhibits metallic character; however, metal functionalization transforms the electronic structure from paramagnetic metal to paramagnetic semiconductor. Bader charge analysis quantifies this transition, revealing charge transfer (Δq) values of 3.377 e, 3.269 e, and 3.299 e for 4Li@$BeN_2$:2$V_{Be}$, 4Na@$BeN_2$:2$V_{Be}$, and 4K@$BeN_2$:2$V_{Be}$, respectively. The greatest transfer occurs in the case of Li atoms, which is likely to be attributed to the polarizability characteristics (i.e., $\alpha^{Li}$ = 24.3, $\alpha^{Na}$ = 24.1, and $\alpha^{K}$ = 43.4).

Figure 5 shows the charge density difference (CDD) maps of 4Li@$BeN_2$:2$V_{Be}$, 4Na@$BeN_2$:2$V_{Be}$, and 4K@$BeN_2$:2$V_{Be}$ monolayers. The yellow (Cyan) color represents the gain (deficit) of charge. The CDD provides rigorous information concerning the nature of chemical bonding, as well as the charge exchange between different species composing the three structures. As far as the electronegativity character is concerned, nitrogen has the highest one (i.e., $\chi^{N}$ = 3.04 > $\chi^{Be}$ = 1.57 Pauling, while $\chi^{Li}$ = 0.98 > $\chi^{Na}$ = 0.93 > $\chi^{K}$ = 0.82 Pauling). Figure 5a (i.e., case of 4Li@$BeN_2$:2$V_{Be}$) shows the depletion of charge from the regions of Be and Li atoms and its accumulation along the chemical bonds and at the nitrogen sites. This trend reveals that the chemical bonds are covalent with partial ionic character. Similar trends are also displayed in Figures 5b and 5c. For these reasons, the PDOS plots in the previous Figure show that the valence band (VB) is mainly attributed to N, while the conduction band (CB) is predominantly contributed by Be and metal dopants (Li, Na, K). The charge transfer is shown in Figure 5 to occur from these latter electropositive atoms to nearby chemical bonds and to N atoms.

**d. $H_2$ Adsorption Mechanism**

To study the $H_2$ uptake capacity of 4Li@$BeN_2$:$2V_{Be}$, 4Na@$BeN_2$:$2V_{Be}$, and 4K@$BeN_2$:$2V_{Be}$, we performed hydrogenation by adding $4H_2$ molecules at a time (i.e., one molecule adsorbed per dopant). Figure 6 shows the top and side views of the optimized configurations of $20H_2$ molecules on 4Li@$BeN_2$:$2V_{Be}$, 4Na@$BeN_2$:$2V_{Be}$, and 4K@$BeN_2$:$2V_{Be}$. The adsorbed $H_2$ molecules begin to form two layers (clouds). The low-lying layer has a stronger $E_{ads}$, whereas the overlying layer is less strongly attached to the substrate. Yet, for instance, in the case of 4Li@$BeN_2$:$2V_{Be}$, it appears capable of absorbing additional $H_2$ molecules. Figure 7a shows the absolute value of the average $E_{ads}$ versus the total number "N" of adsorbed $H_2$ molecules. As mentioned before, throughout the hydrogenation, the recursive step is set to $\Delta N = 4$ molecules, and N is varied up to 20 molecules. Two trends are clearly illustrated in Figure 7a. **(i)** The average $E_{ads}$ decreases with the number of $H_2$ molecules added to the system. Naturally, subsequent groups of $4H_2$ occupy the lowest-energy locations available in the system, and so on. **(ii)** As a threshold/critical energy, we have set the value $|E_{ads}^{crit}|$ = 0.15 eV per molecule, which is broadly used in literature, such as in our recent work [34-35]. So, the absolute values of the average adsorption energy in the case of the adsorption of $20H_2$ molecules are $|E_{ads}|$ = 0.182 eV, 0.191 eV, and 0.177 eV per molecule, being higher than $|E_{ads}^{crit}|$.

Based upon the maximum number (N) of adsorbed $H_2$ molecules, the theoretical gravimetric capacity ($C_T$) can be calculated using equation (3). The calculated theoretical gravimetric capacities on the candidate samples are shown in Figure 7b and have values of 11.64, 9.82, and 8.49 wt%, respectively. These values exceed the DOE ultimate target of 6.50 wt%.

**e. Thermodynamic Analysis**

The calculations of theoretical gravimetric capacity were carried out at 0 K under the assumption of validity of the Born-Oppenheimer "frozen-lattice" approximation. More realistic predictions of the gravimetric capacity could be attempted using the thermodynamic analysis based upon the Langmuir adsorption model [36-37], see equations (4-6) in section 2.

Figure 8 shows the average number of $H_2$ molecules ($N_{ave}$) per primitive cell of 4M@$BeN_2$:$2V_{Be}$ primitive cell that can be adsorbed at given pressure (P) and temperature (T) values (i.e., N-P-T phase diagram). It should be emphasized that the $H_2$ molecules can be stored at low temperature and high pressure and can be released or desorbed at high temperature and low pressure. In practice, the P and T of adsorption are about 30 atm and 25 $^{o}$C, respectively, whereas those of desorption are 3 atm and 100 $^{o}$C, respectively [38]. Table 2 shows the calculated theoretical gravimetric capacity at 0 K ($C_T$) and the effective gravimetric capacity ($C_E$) at the practical adsorption/desorption conditions. It also shows the number of adsorbed $H_2$ molecules ($N_T$), obtained from DFT simulations at 0 K and used to calculate $C_T$. The thermodynamic analysis is employed to estimate the number of $H_2$ molecules under adsorption and desorption conditions, denoted as $N_a$ and $N_d$, respectively, using equation (6). The difference between $N_a$ and $N_d$ represents the practical number of reversibly stored or released $H_2$ molecules ($N_p$), from which the effective storage capacity $C_E$ is calculated using equation (3).

The obtained theoretical gravimetric capacities for 4Li@$BeN_2$:$2V_{Be}$, 4Na@$BeN_2$:$2V_{Be}$, and 4K@$BeN_2$:$2V_{Be}$ are 11.64, 9.82, and 8.49 wt%, respectively. The thermodynamic analysis has shown that under practical adsorption/desorption conditions, these values are reduced to 7.69, 6.37, and 5.13 wt%, respectively. The Li-decoration of $BeN_2$:$2V_{Be}$ persists in maintaining effective gravimetric capacities higher than the required threshold. So, this monolayer should be promising for $H_2$ storage applications.

Table 3 shows the results of our three $H_2$ storage systems and other competitor 2D materials, namely, the theoretical number of adsorbed $H_2$ molecules per simulation cell ($N_T$), the average adsorption energy per $H_2$ molecule ($E_{ads}$), and the theoretical $H_2$ gravimetric capacity ($C_T$). Compared with other computational studies on various 2D systems, our results are competitive and indicate that LM@$BeN_2$:$2V_{Be}$ ML is a promising candidate material for $H_2$ storage applications.

Finally, it is worth mentioning that although beryllium's inherent toxicity precludes direct practical application of $BeN_2$ monolayers, this constraint does not diminish their scientific significance. Rather, $BeN_2$ serves as an invaluable benchmark system, elucidating the fundamental design principles required to achieve exceptional $H_2$ storage performance, specifically, the synergistic combination of ultrahigh gravimetric capacity with thermodynamically optimal adsorption energetics. By establishing these design principles through $BeN_2$, we create a clear roadmap for developing safer, next-generation materials that can replicate or exceed its $H_2$ storage capabilities for fuel cell applications.

## IV. Conclusion

This study employed state-of-the-art computational methods, DFT, AIMD, and thermodynamic analysis, to evaluate the hydrogen storage potential of $BeN_2$ and $BeN_2$:$2V_{Be}$ monolayers decorated with Li, Na, and K. We demonstrate that embedding up to four light metal dopants per primitive cell is thermodynamically stable, as confirmed by binding energies exceeding respective bulk cohesive energies and AIMD simulations at 400 K exhibiting energy fluctuations below 0.50 eV (<1%). Metal decoration induces a transition in the monolayer's electronic structure from metallic to semiconducting character. Bader charge analysis and Charge Density Difference (CDD) plots reveal substantial charge transfer from Be and metal dopants to N sites and chemical bonds, converting LM atoms into cations that polarize $H_2$ molecules and enhance van der Waals interactions. On average, each metal atom accommodates approximately five $H_2$ molecules, yielding theoretical gravimetric capacities of 11.64 wt% (Li), 9.82 wt% (Na), and 8.49 wt% (K), all exceeding the DOE target. Langmuir isotherm modelling predicts reduced adsorption efficiency under ambient conditions; however, Li-decorated $BeN_2$:$2V_{Be}$ maintains effective capacity above DOE specifications, positioning it as a promising candidate for practical hydrogen storage applications.

**Declaration of Competing Interest**

The authors declare that they have no known competing financial interests or personal relationships that could have appeared to influence the work reported in this paper.

## Acknowledgment

This research was conducted using high-performance computing resources at Khalifa University, New York University Abu Dhabi, United Arab Emirates University, and the University of New England. This research was supported by computational resources provided by the Australian Government through the National Computational Infrastructure (NCI Australia) and the Pawsey Supercomputing Research Centre under NCMAS Merit Allocation. We gratefully acknowledge the contributions of Ryan Brady to this work. We are deeply saddened by his passing and honour their valuable contributions to this research.

## References:

[1] R. Etezadi, S. Shivaramakrishnan, R. Wang, R. Khalighi, L. Zhao, I.G. Eschrich, F. Aquino, M. Ibrahim, C. Tasser, T.T. Tsotsis. Hydrogen storage methods, materials and challenges: Scientific exploration. Ind. Eng. Chem. Res. 65 (2026) 1403-1423.

[2] W. Fang, C. Ding, L. Chen, W. Zhou, J. Wang, K. Huang, R. Zhu, J. Wu, B. Liu, Q. Fang, X. Wang, J. Wang. Review of hydrogen storage technologies and the crucial role of environmentally friendly carriers. Energy & Fuel 38 (2024) 13539-13564.

[3] K.S. Novoselov, A.K. Geim, S.V. Morozov, D. Jiang, S.V. Dubonos, I.V. Grigorieva, A.A. Firsov. Electric field effect in atomically thin carbon films. Science 306 (2004) 666-669.

[4] K.S. Novoselov, D. Jiang, F. Schedin, T.J. Booth, V.V. Khotkevich, S.V. Morozov, A.K. Geim. Two-dimensional atomic crystals. PNAS 102 (2005) 10451-10453.

[5] A.K. Geim, K. S. Novoselov. The rise of graphene. Nature Mater. 6 (2007) 183-191.

[6] W. Othman, M. Fahed, S. Hatim, A. Sherazi, G. Berdiyorov, and N. Tit, "Adsorption of CO2 on Fe-doped graphene nano-ribbons: Investigation of transport properties," J. Phys.: Conf. Ser., 869, 012041, 2017, doi: 10.1088/1742-6596/869/1/012041.

[7] D. Yang, P. Dai, X. Jiang, S.M. Alshehri, T. Ahamad, Y. Bando, X. Wang. Methods of preparation of hexagonal boron nitride nanomaterials. Chem. Mater. 36 (2024) 10008-10053.

[8] A. Splendiani, L. Sun, Y. Zhang, T. Li, J. Kim, C.Y. Chim, G. Galli, F. Wang. Emerging photoluminescence in monolayer MoS2. Nano Lett. 10 (2010) 1271-1275.

[9] A. Shaheen, W. Othman, M. Ali, and N. Tit, "Catalyst-induced gas-sensing selectivity in ZnO nanoribbons: Ab-initio investigation at room temperature," Applied Surface Science, 505, 144602, 2020, doi: 10.1016/j.apsusc.2019.144602.

[10] A. Shaheen, M. Ali, W. Othman, and N. Tit, "Origins of Negative Differential Resistance in N-doped ZnO Nano-ribbons: Ab-initio Investigation," Scientific Reports, 9, 9914, 2019, doi: 10.1038/s41598-019-46335-0.

[11] W. Othman et al., "Selective adsorption of H2 on N-doped ZnO nano-ribbons: First-principle analysis," 2018 5th International Conference on Renewable Energy: Generation and Applications (ICREGA), Al Ain, United Arab Emirates, 2018, pp. 227-231, doi: 10.1109/ICREGA.2018.8337638.

[12] N. Tit, W. Othman, A. Shaheen, and M. Ali, "High selectivity of N-doped ZnO nano-ribbons in detecting H-2, O-2 and CO2 molecules: Effect of negative-differential resistance on gas-sensing," Sensor Actuat B: Chem, 270, 167-178, 2018, doi: 10.1016/j.snb.2018.04.175.

[13] I. Alghoul, W. Othman, I. Abdi, T. Hussain, and N. Tit, "Suitable materials for efficient detection of colorectal cancer biomarkers: acumen from DFT," Results in Physics, 78, 108493, 2025, doi: 10.1016/j.rinp.2025.108493.

[14] Y. Gogotsi, B. Anasori. The rise of Mxenes. ACS Nano 13 (2019) 8491-8494.

[15] F. Schwierz. Graphene transistors. Nat. Nanotechnol. 5 (2010) 487-96.

[16] L. Liao, Y.C. Lin, M. Bao, R. Cheng, J. Bai, Y. Liu, Y. Qu, K.L. Wang, Y. Huang, X. Duan. High-speed graphene transistors with self-aligned nanowire gate. Nature 467 (2010) 305-308.

[17] L. Sponza, H. Amara, C. Attaccalite, S. Latil, T. Galvani, F. Paleari, L. Wirtz, F. Ducastelle. Direct and indirect excitons in boron nitride polymorphs: A story of atomic configuration and electronic correlation. Phys. Rev. B 98 (2018) 125206.

[18] Y. Sun, D. Wang, Z. Shuai. Indirect-to-direct band gap crossover in few-layer transition metal dichalcogenides: A theoretical prediction. J. Phys. Chem. C 120 (2016) 21866-21870.

[19] C. Zhang, Q. Sun. A honeycomb BeN2 sheet with a desirable direct band gap and high carrier mobility. J. Phys. Chem. Lett. 7 (2016) 2664-2670.

[20] S. Ni, J. Jiang, W. Wang, X. Wu, Z. Zhuo, Z. Wang. Beryllium dinitride monolayer: a multifunctional direct bandgap anisotropic semiconductor containing polymeric nitrogen with oxygen reduction catalysis and potassium-ion storage capability. J. Mater. Chem. A 13 (2025) 10214.

[21] J. Kang, L. Zhang, S.H. Wei. A unified understanding of the thickness-dependent bandgap transition in hexagonal two-dimensional semiconductors. J. Phys. Chem. Lett. 7 (2016) 597-602.

[22] E. Mohebbi, M. Masoud, S. Fakhrabadi. Investigation of stability, electronic, optical and mechanical properties of honeycomb BeN2 monolayer: A DFT study. Comput. Theor. Chem. 1226 (2023) 114202.

[23] M.R. Shwin Kishore, R. Varunaa, A. Bayani, K. Larsson. Theoretical investigation on BeN2 monolayer for an efficient bifunctional water splitting catalyst. Sci. Rep. 10 (2020) 21411.

[24] W. Zhou, S. Guo, H. Zeng, S. Zhang. High-performance monolayer BeN2 transistors with ultrahigh on-state current: A DFT coupled with NEGF study. IEEE Trans. Electron. Dev. 69 (2022) 4501-4506.

[25] W. Alfalasi, W. Othman, T. Hussain, N. Tit. Multifunctionality of vacancy-induced boron nitride monolayers for metal-ion battery and hydrogen-storage applications. Appl. Surf. Sci. 685 (2025) 162025.

[26] G. Kresse, J. Furthmuller. Efficient iterative schemes for ab-initio total energy calculations using a plane-wave basis set. Phys. Rev. B 54 (1996) 11169.

[27] J.P. Perdew, K. Burke, M. Ernzerhof. General gradient approximation made simple. Phys. Rev. Lett. 77 (1996) 3865.

[28] S. Grimme, J. Antony, S. Ehrlich, and H. Krieg. A consistent and accurate ab initio parametrization of density functional dispersion correction (DFT-D) for the 94 elements H-Pu. J. Chem. Phys. 132 (2010) 154104.

[29] H.J. Monkhorst, J.D. Pack. Special points for Brillouin-zone integrations. Phys. Rev. B 13 (1976) 5188.

[30] G. Henkelman, A. Arnaldsson, and H. Jónsson. A fast and robust algorithm for Bader decomposition of charge density. Comp. Mater. Sci. 36 (2006) 354-360.

[31] S. Khan, N. Kumar, T. Hussain, N. Tit. Functionalized Hf3C2 and Zr3C2 MXenes for suppression of shuttle effect to enhance the performance of sodium-sulfur batteries. J. Power Sources 580 (2023) 233298.

[32] W. Alfalasi, Y.P. Feng, N. Tit. Enhancement of hydrogen storage using functionalized MoSe2/Graphene monolayer and bilayer systems: DFT study. Int. J. Hydrogen Energy 50 (2024) 1189-1203.

[33] A. Hashemi et al. Ultrahigh capacity hydrogen storage in a Li-decorated two-dimensional C2N layer. J. Mater. Chem. A 5 (2017) 2821-2828.

[34] W. Othman, W. Alfalasi, T. Hussain, N. Tit. Light-metal functionalized boron monoxide monolayers as efficient hydrogen storage material: Insights from DFT simulations. J. Energy Storage 98 (2024) 113014.

[35] W. Othman, I. Alghoul, N. Tit, K.F. Aguey-Zinsou, T. Hussain. Computational characterization of advanced hydrogen storage architecture using transition-metal-functionalized C3N5 monolayers. ACS Appl. Energy Mater. 8 (2025) 11614-11624.

[36] H. Bae, M. Park, B. Jang, Y. Kang, J. Park, H. Lee, H. Chung, C. Chung, S. Hong, Y. Kwon, B.I. Yakobson, H. Lee. High-throughput screening of metal-porphyrin-like graphene for selective capture of carbon dioxide. Sci. Rep. 6 (2016) 21788.

[37] H. Yang, H. Bae, M. Park, S. Lee, K.C. Kim, H. Lee. Fe-Porphyrin-like nanostructures for selective ammonia capture under humid conditions. J. Phys. Chem. C 122 (2018) 2046-2052.

[38] H. Lee, W.I. Choi, J. Ihm. Combinatorial search for optimal hydrogen-storage nanomaterials based on polymers. Phys. Rev. Lett. 97 (2006) 056104.

[39] P. Habibi, T.J.H. Vlugt, P. Dey, O.A. Moultos. Reversible hydrogen storage in metal-decorated honeycomb architecture. Int. J. Hydrogen Energy 47 (2022) 33391-33402

[40] S. P. Kaur, T. Hussain, T. Kaewmaraya, and T. J. D. Kumar, "Reversible hydrogen storage tendency of light-metal (Li/Na/K) decorated carbon nitride ($C_9N_4$) monolayer," *Int J Hydrogen Energ,* vol. 48, no. 67, pp. 26301–26313, 2023, doi: 10.1016/j.ijhydene.2023.03.141

[41] A. Vaidyanathan, P. Mane, V. Wagh, and B. Chakraborty, "Vanadium-decorated 2D polyaramid material for high-capacity hydrogen storage: Insights from DFT simulations," *J Energy Storage,* vol. 78, 2024, doi: 10.1016/j.est.2023.109899

[42] P. Mane, S. P. Kaur, M. Singh, A. Kundu, and B. Chakraborty, "Superior hydrogen storage capacity of Vanadium decorated biphenylene (Bi+V): A DFT study," *Int J Hydrogen Energ,* vol. 48, no. 72, pp. 28076–28090, 2023, doi: 10.1016/j.ijhydene.2023.04.033

[43] Y. L. Liu, L. Ren, Y. He, and H. P. Cheng, "Titanium-decorated graphene for high-capacity hydrogen storage studied by density functional simulations," *J Phys-Condens Mat,* vol. 22, no. 44, 2010, doi: 10.1088/0953-8984/22/44/445301

[44] B. Chakraborty, P. Ray, N. Garg, and S. Banerjee, "High capacity reversible hydrogen storage in titanium doped 2D carbon allotrope Ψ-graphene: Density Functional Theory investigations," *Int J Hydrogen Energ,* vol. 46, no. 5, pp. 4154–4167, 2021, doi: 10.1016/j.ijhydene.2020.10.161

[45] Z. Y. Liu, T. Hussain, A. Karton, and S. Er, "Empowering hydrogen storage properties of haeckelite monolayers via metal atom functionalization," *Applied Surface Science,* vol. 556, 2021, doi: 10.1016/j.apsusc.2021.149709

[46] V. Mahamiya, A. Shukla, N. Garg, and B. Chakraborty, "High-capacity reversible hydrogen storage in scandium decorated holey graphyne: Theoretical perspectives," *Int J Hydrogen Energ,* vol. 47, no. 12, pp. 7870–7883, 2022, doi: 10.1016/j.ijhydene.2021.12.112

[47] B. J. Cid *et al.*, "Enhanced reversible hydrogen storage performance of light metal-decorated boron-doped siligene: A DFT study," *Int J Hydrogen Energ,* vol. 47, no. 97, pp. 41310–41319, 2022, doi: 10.1016/j.ijhydene.2022.03.153

[48] B. J. Cid, A. N. Sosa, A. Miranda, L. A. Pérez, F. Salazar, and M. Cruz-Irisson, "Hydrogen storage on metal decorated pristine siligene and metal decorated boron-doped siligene," *Mater Lett,* vol. 293, 2021, doi: 10.1016/j.matlet.2021.129743

[49] A. Hashmi, M. U. Farooq, I. Khan, J. Son, and J. Hong, "Ultra-high capacity hydrogen storage in a Li decorated two-dimensional $C_2N$ layer," *J Mater Chem A,* vol. 5, no. 6, pp. 2821–2828, 2017, doi: 10.1039/c6ta08924k

[50] L. F. Z. Wang, X. F. Chen, H. Y. Du, Y. Q. Yuan, H. Qu, and M. Zou, "First-principles investigation on hydrogen storage performance of Li, Na and K decorated borophene," *Applied Surface Science,* vol. 427, pp. 1030–1037, 2018, doi: 10.1016/j.apsusc.2017.08.126

[51] Y. H. Guo *et al.*, "A comparative study of the reversible hydrogen storage behavior in several metal decorated graphyne," *Int J Hydrogen Energ,* vol. 38, no. 10, pp. 3987–3993, 2013, doi: 10.1016/j.ijhydene.2013.01.064

[52] Y. Gao, H. N. Zhang, H. Z. Pan, Q. F. Li, and J. J. Zhao, "Ultrahigh hydrogen storage capacity of holey graphyne," *Nanotechnology,* vol. 32, no. 21, 2021, doi: 10.1088/1361-6528/abe48d

[53] P. Panigrahi *et al.*, "Selective decoration of nitrogenated holey graphene ($C_2N$) with titanium clusters for enhanced hydrogen storage application," *Int J Hydrogen Energ,* vol. 46, no. 10, pp. 7371–7380, 2021, doi: 10.1016/j.ijhydene.2020.11.222

[54] P. Mane, S. P. Kaur, and B. Chakraborty, "Enhanced reversible hydrogen storage efficiency of zirconium-decorated biphenylene monolayer: A computational study," *Energy Storage,* vol. 4, no. 6, 2022, doi: 10.1002/est2.377

[55] P. A. Denis and F. Iribarne, "Hydrogen storage in doped biphenylene based sheets," *Comput Theor Chem,* vol. 1062, pp. 30–35, 2015, doi: 10.1016/j.comptc.2015.03.012

[56] V. Mahamiya, A. Shukla, and B. Chakraborty, "Ultrahigh reversible hydrogen storage in K and Ca decorated 4-6-8 biphenylene sheet," *Int J Hydrogen Energ,* vol. 47, no. 99, pp. 41833–41847, 2022, doi: 10.1016/j.ijhydene.2022.01.216

[57] Z. Y. Liu, S. Liu, and S. Er, "Hydrogen storage properties of Li-decorated $B_2S$ monolayers: A DFT study," *Int J Hydrogen Energ,* vol. 44, no. 31, pp. 16803–16810, 2019, doi: 10.1016/j.ijhydene.2019.04.234

[58] T. W. Wang and Z. Y. Tian, "Yttrium-decorated $C_{48}B_{12}$ as hydrogen storage media: A DFT study," *Int J Hydrogen Energ,* vol. 45, no. 46, pp. 24895–24901, 2020, doi: 10.1016/j.ijhydene.2020.02.025

[59] M. Dixit, T. A. Maark, and S. Pal, "Ab initio and periodic DFT investigation of hydrogen storage on light metal-decorated MOF-5," *Int J Hydrogen Energ,* vol. 36, no. 17, pp. 10816–10827, 2011, doi: 10.1016/j.ijhydene.2011.05.165

[60] A. Kundu and B. Chakraborty, "Yttrium doped covalent triazine frameworks as promising reversible hydrogen storage material: DFT investigations," *Int J Hydrogen Energ,* vol. 47, no. 71, pp. 30567–30579, 2022, doi: 10.1016/j.ijhydene.2022.06.315

[61] B. Chakraborty, P. Mane, and A. Vaidyanathan, "Hydrogen storage in scandium decorated triazine based g-$C_3N_4$: Insights from DFT simulations," *Int J Hydrogen Energ,* vol. 47, no. 99, pp. 41878–41890, 2022, doi: 10.1016/j.ijhydene.2022.02.185

[62] L. H. Yuan *et al.*, "Hydrogen storage capacity on Ti-decorated porous graphene: First-principles investigation," *Applied Surface Science,* vol. 434, pp. 843–849, 2018, doi: 10.1016/j.apsusc.2017.10.231

[63] P. Mane, A. Vaidyanathan, and B. Chakraborty, "Graphitic carbon nitride (g-$C_3N_4$) decorated with Yttrium as potential hydrogen storage material: Acumen from quantum simulations," *Int J Hydrogen Energ,* vol. 47, no. 99, pp. 41898–41910, 2022, doi: 10.1016/j.ijhydene.2022.04.184

[64] S. Dong *et al.*, "Construction of transition metal-decorated boron doped twin-graphene for hydrogen storage: A theoretical prediction," *Fuel,* vol. 304, 2021, doi: 10.1016/j.fuel.2021.121351

[65] X. Y. Liang, S. P. Ng, N. Ding, and C. M. L. Wu, "Strain-induced switch for hydrogen storage in cobalt-decorated nitrogen-doped graphene," *Applied Surface Science,* vol. 473, pp. 174–181, 2019, doi: 10.1016/j.apsusc.2018.12.132

[66] L. H. Yuan *et al.*, "First-principles study of V-decorated porous graphene for hydrogen storage," *Chem Phys Lett,* vol. 726, pp. 57–61, 2019, doi: 10.1016/j.cplett.2019.04.026

[67] S. B. Chu, L. B. Hu, X. R. Hu, M. K. Yang, and J. B. Deng, "Titanium-embedded graphene as high-capacity hydrogen-storage media," *Int J Hydrogen Energ,* vol. 36, no. 19, pp. 12324–12328, 2011, doi: 10.1016/j.ijhydene.2011.07.015

[68] A. N. Sosa *et al.*, "Light metal functionalized two-dimensional siligene for high capacity hydrogen storage: DFT study," *Int J Hydrogen Energ,* vol. 46, no. 57, pp. 29348–29360, 2021, doi: 10.1016/j.ijhydene.2020.10.175

[69] D. Q. Lin *et al.*, "Potassium-doped $PC_{71}BM$ for hydrogen storage: Photoelectron spectroscopy and first-principles studies," *Int J Hydrogen Energ,* vol. 46, no. 24, pp. 13061–13069, 2021, doi: 10.1016/j.ijhydene.2021.01.061

[70] E. Eroglu, S. Aydin, and M. Simsek, "Effect of boron substitution on hydrogen storage in Ca/DCV graphene: A first-principle study," *Int J Hydrogen Energ,* vol. 44, no. 50, pp. 27511–27528, 2019, doi: 10.1016/j.ijhydene.2019.08.186

[71] N. Zheng, S. L. Yang, H. X. Xu, Z. G. Lan, Z. Wang, and H. S. Gu, "A DFT study of the enhanced hydrogen storage performance of the Li-decorated graphene nanoribbons," *Vacuum,* vol. 171, 2020, doi: 10.1016/j.vacuum.2019.109011

[72] Q. Wu, M. M. Shi, X. Huang, Z. S. Meng, Y. H. Wang, and Z. H. Yang, "A first-principles study of Li and Na co-decorated $T_{4,4,4}$-graphyne for hydrogen storage," *Int J Hydrogen Energ,* vol. 46, no. 11, pp. 8104–8112, 2021, doi: 10.1016/j.ijhydene.2020.12.016

[73] A. N. Sosa, B. J. Cid, Á. Miranda, L. A. Pérez, G. H. Cocoletzi, and M. Cruz-Irisson, "A DFT investigation: High-capacity hydrogen storage in metal-decorated doped germanene," *J Energy Storage,* vol. 73, p. 108913, 2023, doi: 10.1016/j.est.2023.108913

[74] Q. Yin, G. Bi, R. Wang, Z. Zhao, and K. Ma, "High-capacity hydrogen storage in lithium decorated penta-$BN_2$: A first-principles study," *Journal of Power Sources,* vol. 591, p. 233814, 2024, doi: 10.1016/j.jpowsour.2023.233814

[75] J. B. Hao *et al.*, "An investigation of Li-decorated N-doped penta-graphene for hydrogen storage," *Int J Hydrogen Energ,* vol. 46, no. 50, pp. 25533–25542, 2021, doi: 10.1016/j.ijhydene.2021.05.089

[76] Q. Q. Yin, G. X. Bi, R. K. Wang, Z. H. Zhao, and K. Ma, "High-capacity hydrogen storage in Li-decorated defective penta-$BN_2$: A DFT-D2 study," *Int J Hydrogen Energ,* vol. 48, no. 67, pp. 26288–26300, 2023, doi: 10.1016/j.ijhydene.2023.03.309

[77] Y. L. Yong, S. Hu, Z. J. Zhao, R. L. Gao, H. L. Cui, and Z. L. Lv, "Potential reversible and high-capacity hydrogen storage medium: Li-decorated $B_3S$ monolayers," *Mater Today Commun,* vol. 29, 2021, doi: 10.1016/j.mtcomm.2021.102938

[78] Y. F. Zhang, P. P. Liu, and X. L. Zhu, "Li decorated penta-silicene as a high capacity hydrogen storage material: A density functional theory study," *Int J Hydrogen Energ,* vol. 46, no. 5, pp. 4188–4200, 2021, doi: 10.1016/j.ijhydene.2020.10.193

[79] L. Bi *et al.*, "Density functional theory study on hydrogen storage capacity of metal-embedded penta-octa-graphene," *Int J Hydrogen Energ,* vol. 47, no. 76, pp. 32552–32564, 2022, doi: 10.1016/j.ijhydene.2022.07.134

[80] M. Shams and A. Reisi-Vanani, "Potassium decorated γ-graphyne as hydrogen storage medium: Structural and electronic properties," *Int J Hydrogen Energ,* vol. 44, no. 10, pp. 4907–4918, 2019, doi: 10.1016/j.ijhydene.2019.01.010

[81] A. L. Marcos-Viquez, A. Miranda, M. Cruz-Irisson, and L. A. Pérez, "Tin carbide monolayers decorated with alkali metal atoms for hydrogen storage," *Int J Hydrogen Energ,* vol. 47, no. 97, pp. 41329–41335, 2022, doi: 10.1016/j.ijhydene.2021.12.204

[82] S. Haldar, S. Mukherjee, and C. V. Singh, "Hydrogen storage in Li, Na and Ca decorated and defective borophene: a first principles study," *Rsc Advances,* vol. 8, no. 37, pp. 20748–20757, 2018, doi: 10.1039/c7ra12512g

[83] N. Khossossi *et al.*, "Hydrogen storage characteristics of Li and Na decorated 2D boron phosphide," *Sustain Energ Fuels,* vol. 4, no. 9, pp. 4538–4546, 2020, doi: 10.1039/d0se00709a

[84] T. Kaewmaraya *et al.*, "Ultrahigh hydrogen storage using metal-decorated defected biphenylene," *Applied Surface Science,* vol. 629, p. 157391, 2023, doi: 10.1016/j.apsusc.2023.157391

[85] P. Habibi, T. J. H. Vlugt, P. Dey, and O. A. Moultos, "Reversible Hydrogen Storage in Metal-Decorated Honeycomb Borophene Oxide," *Acs Applied Materials & Interfaces,* vol. 13, no. 36, pp. 43233–43240, 2021, doi: 10.1021/acsami.1c09865

[86] S. H. Lu, S. M. Zhang, and X. J. Hu, "Ab Initio Study of High-Capacity Hydrogen Storage in Lithium-Shrouded Honeycomb Borophene Oxide Nanosheet," *J Phys Chem C,* vol. 126, no. 49, pp. 20762–20772, 2022, doi: 10.1021/acs.jpcc.2c06621

[87] Y. H. Wei, F. Gao, J. G. Du, and G. Jiang, "Hydrogen storage on Li-decorated BN: a first-principle calculation insight," *J Phys D Appl Phys,* vol. 54, no. 44, 2021, doi: 10.1088/1361-6463/ac0fab

# Figures

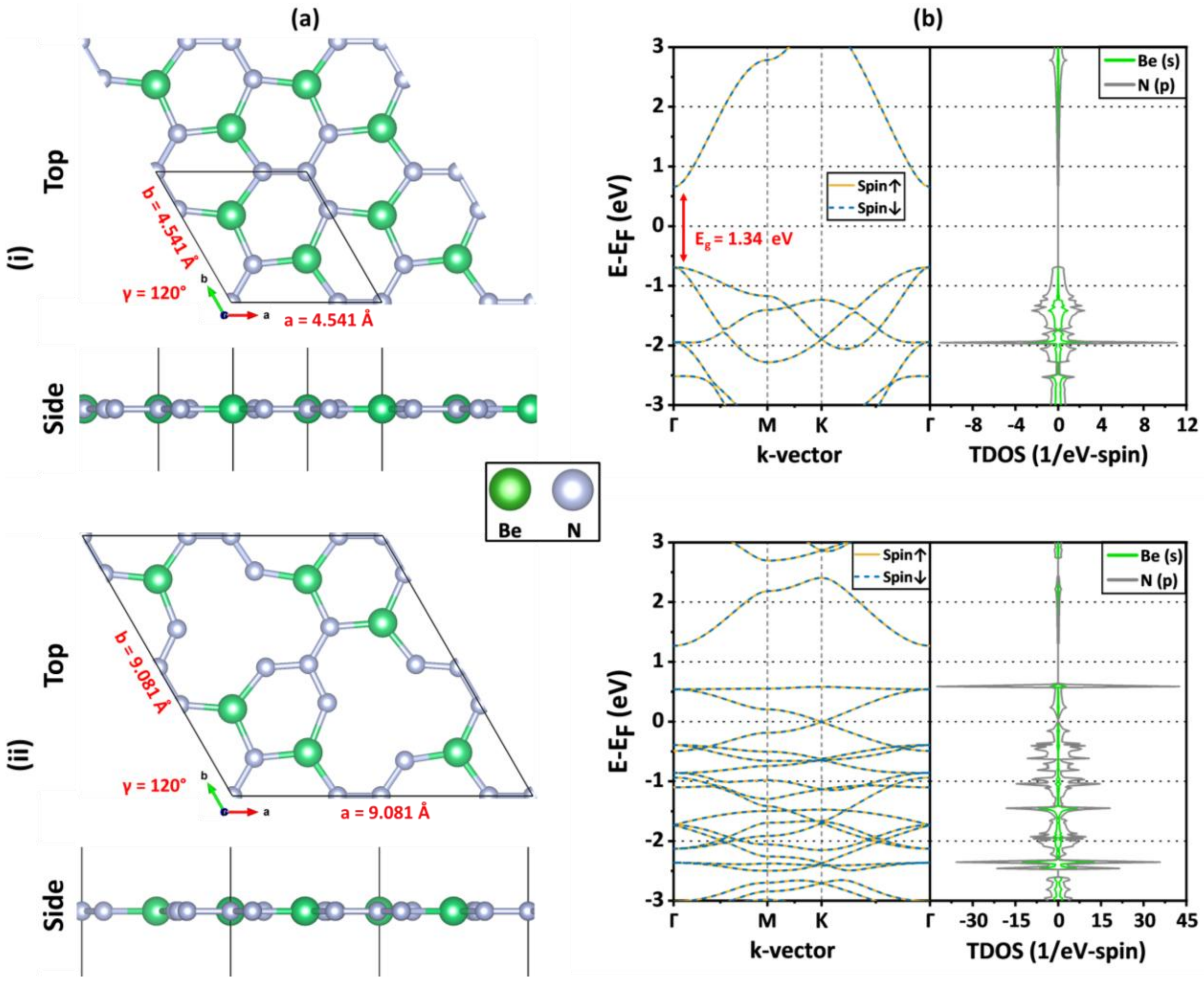

**Figure 1.** (a) Optimized atomic structures (top and side views) and (b) spin-polarized electronic band structure with corresponding projected/total density of states (PDOS/TDOS) of (i) pristine $BeN_2$ and (ii) vacancy-defected $BeN_2$ ($BeN_2$:$2V_{Be}$). The primitive cell of $BeN_2$ contains 2 Be and 4 N atoms arranged in a hexagonal lattice with lattice parameters a = b = 4.541 Å and c = 20 Å (vacuum spacing), α = β = 90°, and γ = 120°. The 2×2 supercell containing two Be vacancies ($BeN_2$:$2V_{Be}$) consists of 6 Be and 16 N atoms with lattice parameters a = b = 9.081 Å. Pristine $BeN_2$ exhibits a direct band gap of 1.34 eV with a non-magnetic ground state, whereas vacancy engineering induces metallic behavior while preserving the non-magnetic character. The high-symmetry M, K, and Γ points correspond to 0 0.7989, 1.2601, 2.1826 $Å^{-1}$ for $BeN_2$, and 0.3994, 0.6301, and 1.0913 $Å^{-1}$ for $BeN_2$:$2V_{Be}$. The Fermi energy ($E_F$) is set to zero.

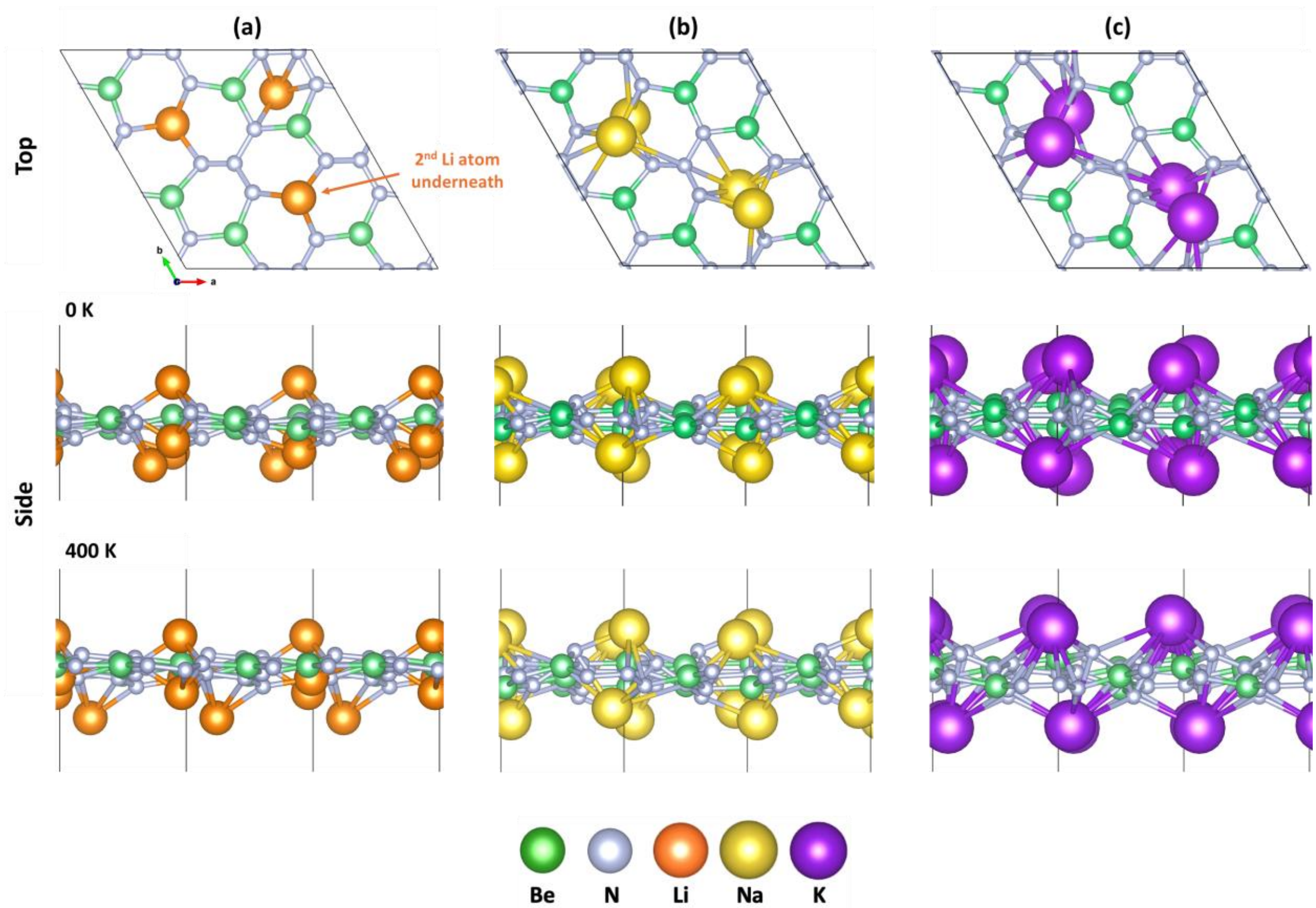

**Figure 2.** Optimized structures (top and side views) of (a) $4Li@BeN_2{:}2V_{Be}$, (b) $4Na@BeN_2{:}2V_{Be}$, and (c) $4K@BeN_2{:}2V_{Be}$ monolayers. AIMD simulations at 400 K confirm the thermal stability of the light-metal-functionalized $BeN_2{:}2V_{Be}$ monolayers.

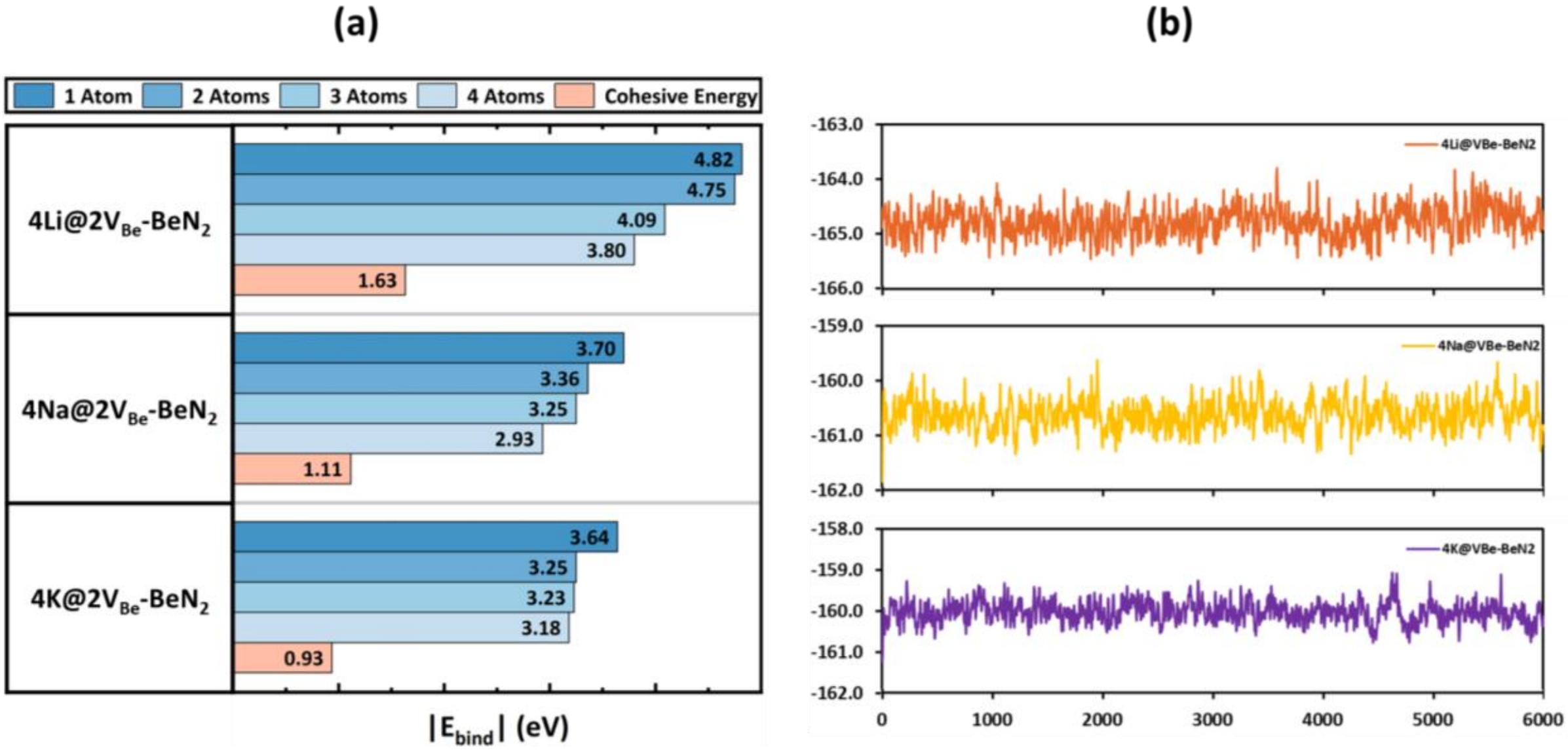

**Figure 3.** (a) Average binding energy ($|E_{bind}|$) per light-metal atom doped to $BeN_2$:$2V_{Be}$ monolayers (four metal atoms per 2×2 supercell), compared with the corresponding cohesive energy of the bulk metals. (b) Ab initio molecular dynamics (AIMD) simulations 4M@$BeN_2$:$2V_{Be}$ (M = Li, Na, K), demonstrating structural stability and absence of metal clustering.

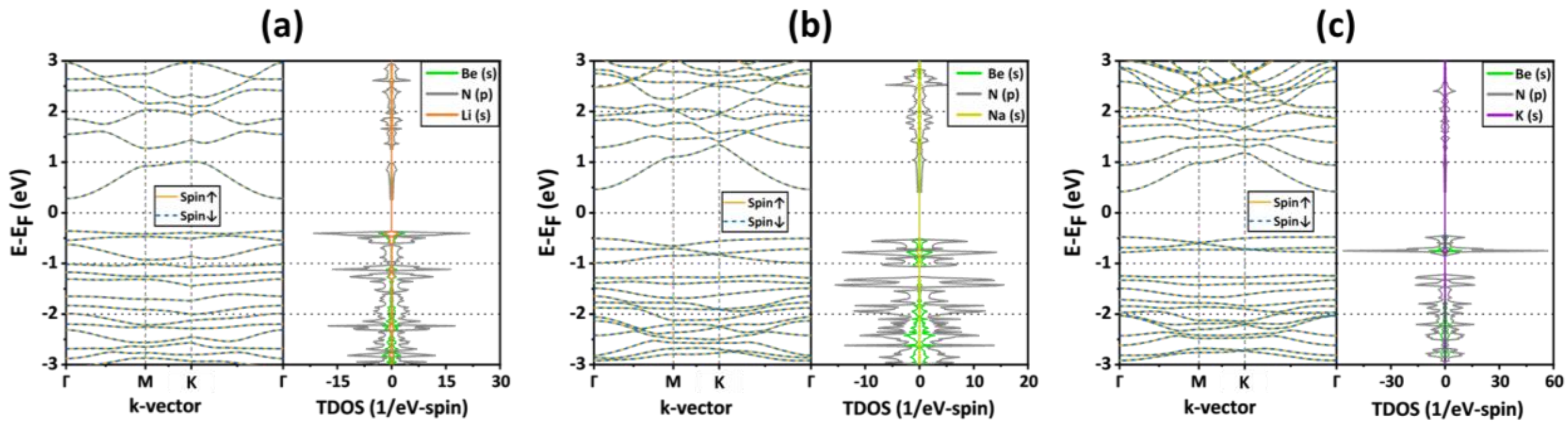

**Figure 4.** Spin-polarized electronic band structures and projected density of states (PDOS) of (a) 4Li@$BeN_2$:$2V_{Be}$, (b) 4Na@$BeN_2$:$2V_{Be}$, and (c) 4K@$BeN_2$:$2V_{Be}$ monolayers. The high-symmetry M, K, and Γ points correspond to 0.3994, 0.6301, and 1.0913 $Å^{-1}$, respectively. The Fermi energy ($E_F$) is set to zero.

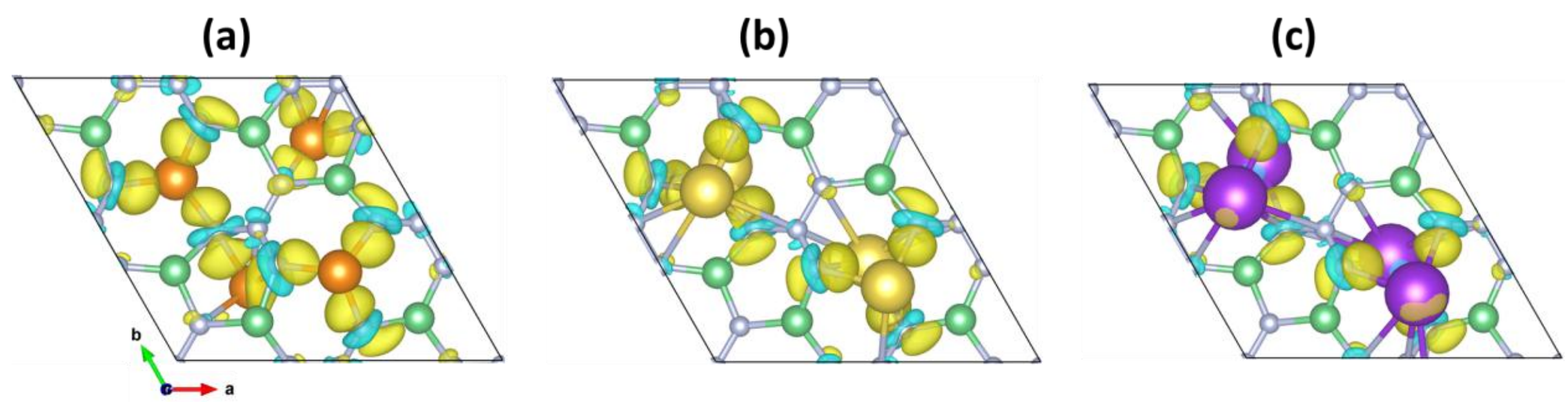

**Figure 5.** Charge density difference (CDD) maps showing the bonding characteristics of (a) $4Li@BeN_2:2V_{Be}$, (b) $4Na@BeN_2:2V_{Be}$, and (c) $4K@BeN_2:2V_{Be}$ monolayers. Yellow and cyan isosurfaces denote regions of charge accumulation and depletion, respectively (isosurface value: 0.005 e/Bohr$^3$).

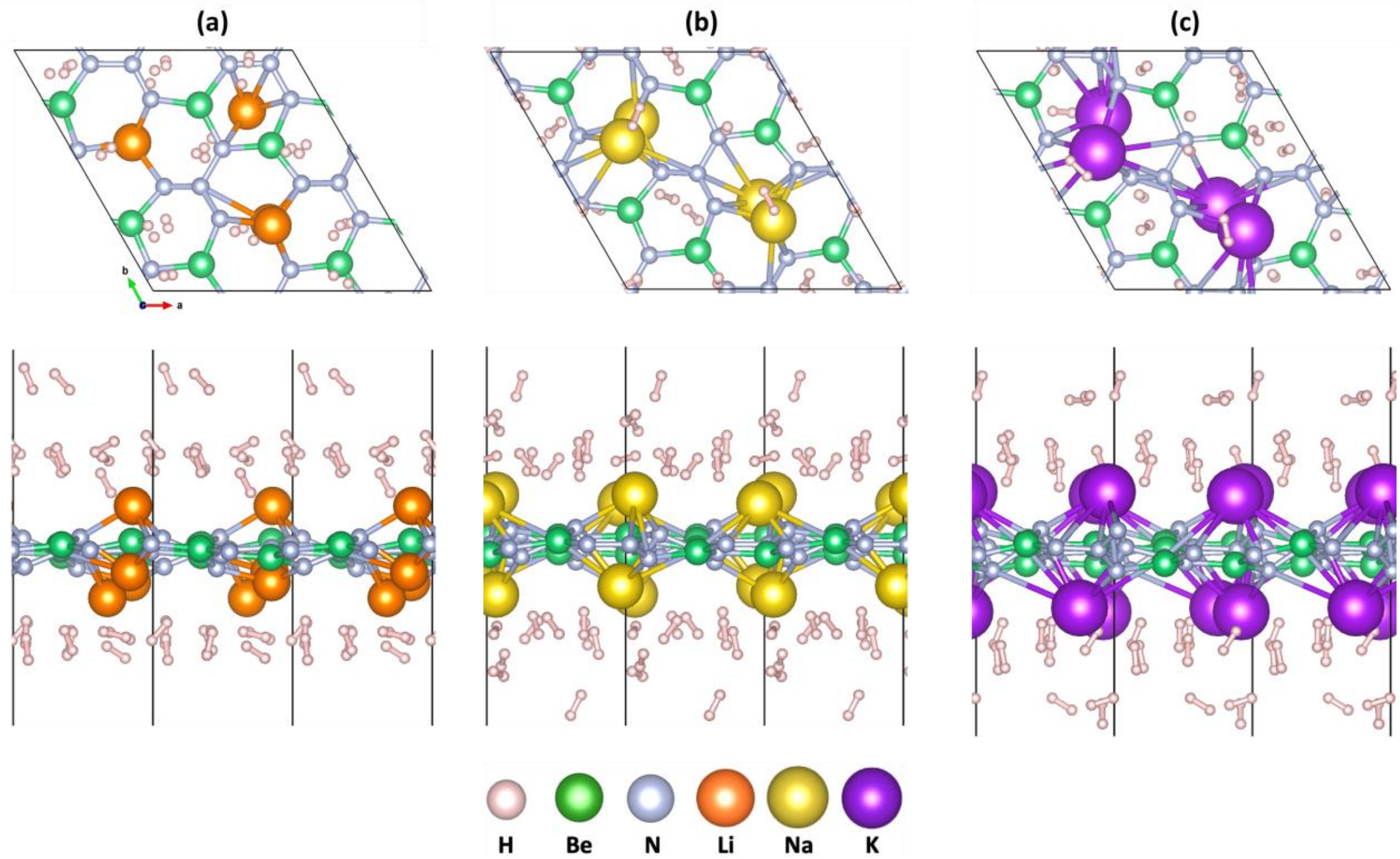

**Figure 6.** Optimized atomic structures (top and side views) of fully hydrogenated 4M@$BeN_2$:2$V_{Be}$ monolayers (M = Li, Na, K). (a) 4Li@$BeN_2$:2$V_{Be}$ with 20 $H_2$ molecules, (b) 4Na@$BeN_2$:2$V_{Be}$ with 20 $H_2$ molecules, and (c) 4K@$BeN_2$:2$V_{Be}$ with 20 $H_2$ molecules.

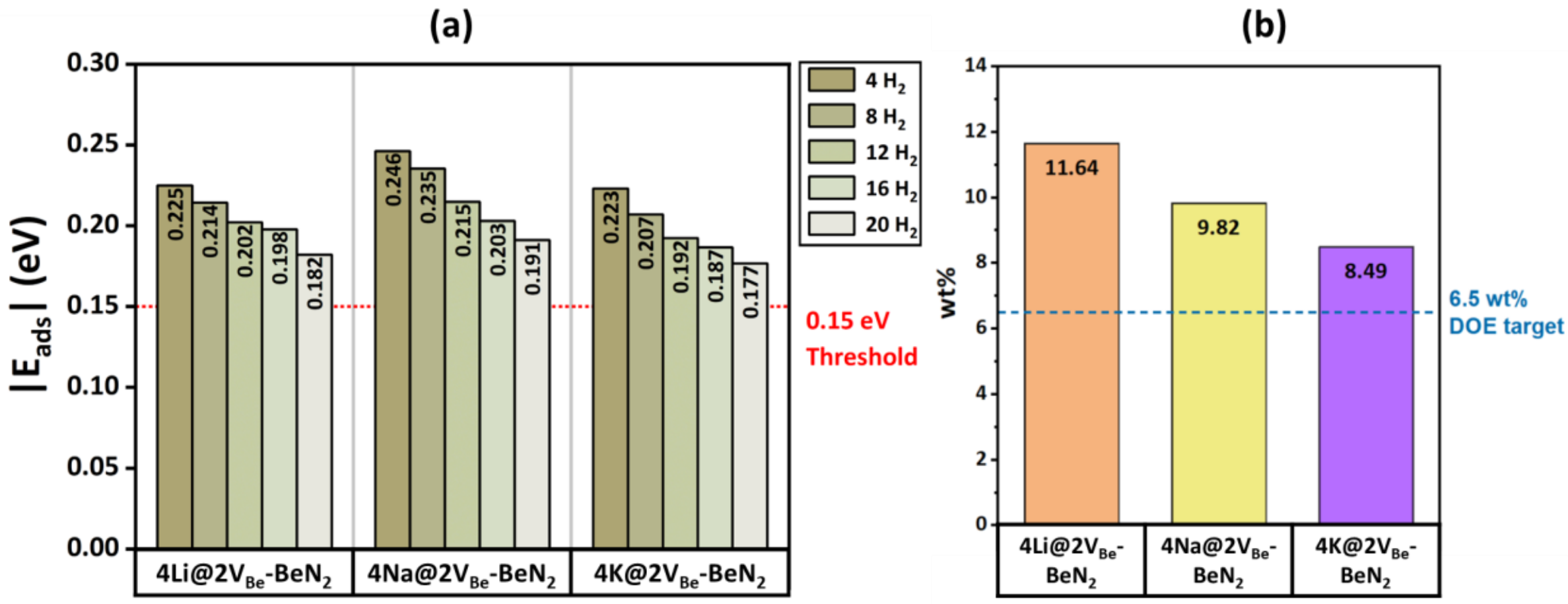

**Figure 7.** $H_2$ storage performance of fully hydrogenated 4M@$BeN_2$:$2V_{Be}$ monolayers (M = Li, Na, K). (a) Average adsorption energy ($E_{ads}$) per $H_2$ molecule within the optimal thermodynamic window for reversible hydrogen storage (~0.15–0.25 eV). (b) Corresponding theoretical gravimetric hydrogen storage capacities (wt%) under maximum hydrogenation exceeding the DOE ultimate target of 6.5 wt%.

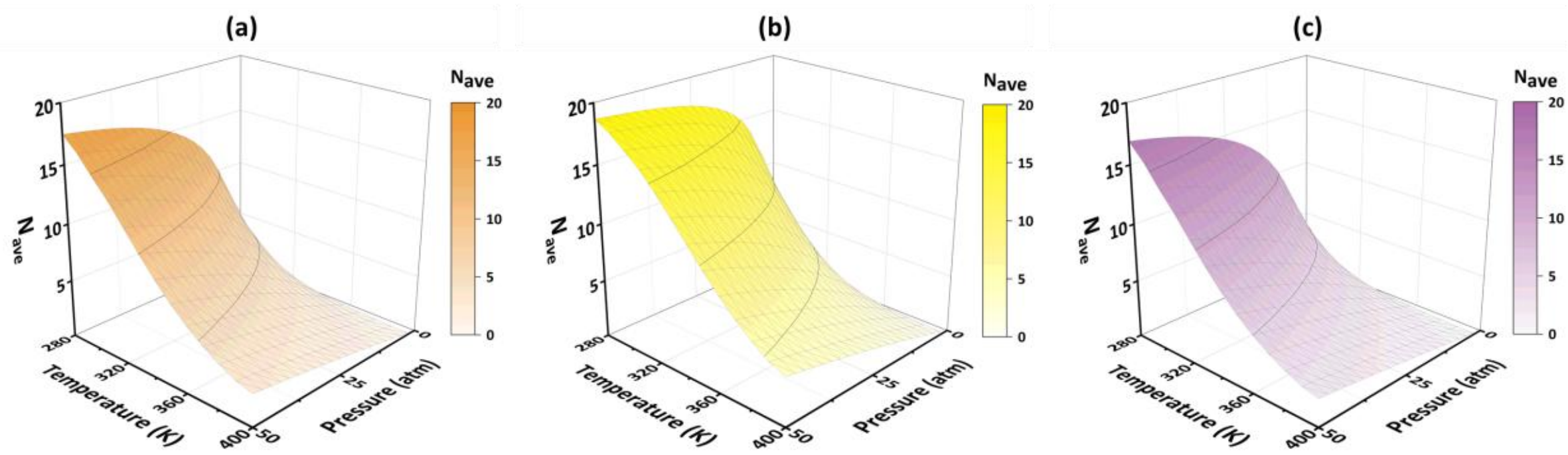

**Figure 8.** Thermodynamic adsorption profiles illustrating the three-dimensional variation of the average number of adsorbed $H_2$ molecules (N-avg) as a function of temperature and pressure for (a) $4Li@BeN_2:2V_{Be}$, (b) $4Na@BeN_2:2V_{Be}$, and (c) $4K@BeN_2:2V_{Be}$.

**Table 1.** Electronic properties of pristine $BeN_2$, $BeN_2$:$2V_{Be}$, and 4(M)@$BeN_2$:$2V_{Be}$ (M = Li, Na, K) monolayers. Δq represents the Bader charge transfer between metal atoms and the substrate, $E_F$ corresponds to the Fermi level, $E_g$ denotes the band gap, while **M** refers to the total magnetization.

| System | Δq (e/MI) | $E_F$ (eV) | $E_g$(eV) | M ($\mu_B$) |
|---|---|---|---|---|
| Pristine $BeN_2$ | NA | -3.74 | 1.34 (↑)(↓) | 0 |
| $2V_{Be}$-$BeN_2$ | NA | -4.37 | 0.02 (↑)(↓) | 0 |
| 4Li@$2V_{Be}$-$BeN_2$ | 3.377 | -2.10 | 0.64 (↑)(↓) | 0 |
| 4Na@$2V_{Be}$-$BeN_2$ | 3.269 | -1.30 | 0.97 (↑)(↓) | 0 |
| 4Na@$2V_{Be}$-$BeN_2$ | 3.299 | -0.75 | 0.90 (↑)(↓) | 0 |

**Table 2.** Theoretical and practical $H_2$ storage performance of light-metal-functionalized $BeN_2$:$2V_{Be}$ monolayers. The theoretical maximum number of adsorbed $H_2$ molecules ($N_T$), obtained from DFT simulations, is used to calculate the theoretical gravimetric storage capacities ($C_T$) according to Eq. (3). Thermodynamic analysis using Eq. (6) estimates the number of adsorbed $H_2$ molecules under practical adsorption (P = 30 atm and T = 25 °C) and desorption conditions (3 atm and T = 100 °C), denoted as $N_a$ and $N_d$, respectively. The difference ($N_p = N_a - N_d$) represents the reversibly stored $H_2$ molecules, from which the effective gravimetric storage capacity ($C_E$) is calculated.

| Storage material | $C_T$ (wt%) | $N_T$ (molecule) | $N_a$ (molecule) | $N_d$ (molecule) | $N_p$ (molecule) | $C_E$ (wt%) |
|---|---|---|---|---|---|---|
| 4Li@2$V_{Be}$-$BeN_2$ | 11.64 | 20 | 12.98 | 0.34 | 12.64 | 7.69 |
| 4Na@2$V_{Be}$-$BeN_2$ | 9.82 | 20 | 15.68 | 0.56 | 15.12 | 6.37 |
| 4Na@2$V_{Be}$-$BeN_2$ | 8.49 | 20 | 11.94 | 0.28 | 11.66 | 5.13 |

**Table 3.** Benchmark comparison of Light-metal-decorated $BeN_2$:$2V_{Be}$ with previously reported two-dimensional hydrogen storage materials based on the theoretical maximum number of adsorbed $H_2$ molecules per simulation cell ($N_T$), the average adsorption energy per $H_2$ molecule ($E_{ads}$), and the corresponding theoretical gravimetric hydrogen storage capacity ($C_T$). The computational methodologies adopted in each referenced study are indicated.

| Hydrogen Storage System | $N_T$ | $E_{ads}$ (eV) | $C_T$ (wt%) |
|---|---|---|---|
| Li-decorated $2V_{Be}$-$BeN_2$* (this work) | 20 | -0.18 | 11.64 |
| Na-decorated $2V_{Be}$-$BeN_2$* (this work) | 20 | -0.19 | 9.82 |
| K-decorated $2V_{Be}$-$BeN_2$* (this work) | 20 | -0.18 | 8.49 |
| Li-decorated $C_9N_4$* [40] | 6 | -0.20 | 11.90 |
| V-decorated 2DPA-I** [41] | 7 | -0.44 | 7.29 |
| V-decorated biphenylene* [42] | 7 | -0.47 | 10.30 |
| Ti-decorated graphene** [43] | 8 | -0.42 | 7.80 |
| Ti-doped ψ-graphene* [44] | 9 | -0.30 | 13.41 |
| Li-decorated B@$r_{57}$haeckelite* [45] | 12 | -0.16 | 10.00 |
| Sc-doped Holey graphyne*** [46] | 5 | -0.36 | 9.80 |
| Ca-decorated B-doped siligene** [47] | 7 | -0.20 | 13.79 |
| Li-decorated B-doped siligene** [48] | 4 | -0.17 | 12.71 |
| Li-decorated $C_2N$** [49] | 20 | -0.14 | 13.00 |
| Li-decorated borophene** [50] | 14 | -0.20 | 13.96 |
| Li-decorated graphyne** [51] | 12 | -0.26 | 18.60 |
| Li-decorated Holey graphyne** [52] | 24 | -0.22 | 12.80 |
| Ti-decorated $C_2N$* [53] | 10 | -0.28 | 6.80 |
| Zr-decorated biphenylene* [54] | 9 | -0.40 | 9.95 |
| Li-decorated biphenylene***** [55] | 12 | -0.20 | 7.40 |
| K-decorated biphenylene** [56] | 5 | -0.24 | 11.90 |
| Li-decorated $B_2S$ honeycomb* [57] | 12 | -0.14 | 9.10 |
| Y-decorated $C_{48}B_{12}$** [58] | 72 | -0.46 | 7.51 |
| Li-decorated MOF-5** [59] | 18 | _ | 4.30 |
| Y-decorated covalent triazine frameworks* [60] | 7 | -0.33 | 7.30 |
| Sc-decorated g-$C_3N_4$** [61] | 7 | -0.39 | 8.55 |
| Ti-decorated graphene**** [62] | 8 | -0.46 | 6.11 |
| Y-decorated g-$C_3N_4$** [63] | 9 | -0.33 | 8.55 |
| Ti-decorated B-doped twin-graphene* [64] | 8 | -0.20 | 4.95 |
| Co-decorated N-doped graphene** [65] | 28 | -0.19 | 11.36 |
| V-decorated porous graphene**** [66] | 6 | -0.56 | 4.58 |
| Ti-decorated graphene** [67] | 8 | -0.21 | 6.30 |
| Na-decorated siligene** [68] | 32 | -0.16 | 14.16 |
| K-doped $PC_{71}BM$** [69] | 45 | -0.14 | 6.22 |
| Ca-decorated DCV graphene**** [70] | 14 | -0.10 | 5.80 |
| Li-decorated graphene nanoribbons** [71] | 8 | -0.24 | 3.80 |
| Li-decorated $T_{4,4,4}$-graphyne* [72] | 16 | -0.20 | 10.46 |
| K-decorated Ga-doped germanene** [73] | 36 | -0.19 | 8.19 |
| Li-decorated P-$BN_2$** [74] | 28 | -0.16 | 13.27 |
| Li-decorated N-doped penta-graphene** [75] | 12 | -0.24 | 7.88 |
| Li-decorated defective penta-$BN_2$** [76] | 16 | -0.14 | 9.17 |
| Li-decorated $B_3S$* [77] | 12 | -0.17 | 7.70 |
| Li-decorated penta-silicene**** [78] | 12 | -0.22 | 6.42 |
| Li-decorated penta-octa-graphene* [79] | 3 | -0.22 | 9.90 |
| γ–graphyne**** [80] | 69 | -0.20 | 13.90 |
| K-decorated SnC** [81] | 6 | -0.20 | 5.50 |
| Li-decorated borophene**** [82] | 10 | -0.36 | 9.00 |
| Li-decorated boron phosphide** [83] | 16 | -0.19 | 7.40 |
| Li-decorated defected biphenylene* [84] | 14 | -0.20 | 8.75 |
| Li-decorated honeycomb borophene oxide** [85] | 32 | -0.24 | 8.3 |

| Li-decorated honeycomb borophene oxide** [86] | 8 | -0.22 | 9.84 |
|---|---|---|---|
| Li-decorated $B_4N$** [87] | 16 | -0.16 | 6.23 |

*GGA-PBE/DFT-D3
**GGA-PBE/DFT-D2
***GGA-PBE/DFT-D2 and HSE06
****GGA-PBE/DFT-D
*****VDW-DF/DZP